\documentclass[twocolumn,aps,prl,showpacs,groupedaddress]{revtex4-1}

\usepackage{epsfig,amssymb,amsmath,mathrsfs,color} 
\newtheorem{lemma}{Lemma} 
 \newtheorem{theorem}{Theorem}
 \newtheorem{definition}{Definition}
 \def\Proof{\medskip\par\noindent{\bf Proof.
  }}
\def\qed{$\,\blacksquare$\par} 
 \def\>{\rangle}
\def\<{\langle} \def\trnsfrm#1{\mathscr
  #1} \def\rA{{\rm A}}\def\rB{{\rm
    B}}\def\rC{{\rm C}}    \def\rI{{\rm I}}   \def\rX{{\rm X}} \def\rY{{\rm Y}}\def\rZ{{\rm Z}}
  \def\tA{\trnsfrm A}\def\tB{\trnsfrm
  B} 
\def\tI{\trnsfrm I}
  
  \def\Cntset{{\mathfrak E}}
 
\def\Stset{{\mathfrak S}} \def\Trnset{{\mathfrak T}}
  
  \def\Tr{{\rm Tr}}
 \def\Reals{{\mathbb R}}

\begin{document}

\title{Discord and non-classicality in probabilistic theories}
\author{Paolo Perinotti}\email{paolo.perinotti@unipv.it}
\affiliation{{\em QUIT Group}, Dipartimento di Fisica ``A. Volta'',
  via Bassi 6, 27100 Pavia, Italy} \homepage{http://www.qubit.it}
\date{\today}
\begin{abstract}
  Quantum discord quantifies non-classical correlations in quantum
  states. We introduce discord for states in causal probabilistic
  theories, inspired by the original definition proposed in Ref.
  \cite{ollzur}. We show that the only probabilistic theory in which
  all states have null discord is classical probability theory.
  Non-null discord is then not just a quantum feature, but a generic
  signature of non-classicality.
\end{abstract}
\pacs{03.67.-a, 03.67.Ac, 03.65.Ta}\maketitle

Non-locality plays a crucial role in quantum foundations. Entanglement
is indeed the source of the most striking quantum paradoxes, since
Schr\"odinger's cat paper \cite{schrocat}, or the incompleteness
argument by Einstein, Podolsky and Rosen \cite{epr}.  Any attempt to
retain locality of physical properties is doomed to give up a
realistic interpretation as proved by Bell's inequality argument
\cite{bell,chsh}. Additional arguments against the existence of local
realistic theories compatible with the quantum statistics were later
proved \cite{mermin,ghz,hardy,cabello}.  The approach to the
discussion on non-classical aspects of quantum theory, such as
entanglement and non-locality, radically changed in recent years, due
to the increasing interest in information processing within general
probabilistic theories \cite{barrett,mauro,purif,barn,short,long} as a
new viewpoint for looking at quantum and classical theories from the
outside. The notion of {\em entanglement} can be easily extended to
general probabilistic theories, being just the negation of the
separability property. From this point of view, classical theory is
not the only one forbidding entangled states (see e.g.
\cite{bitcomm}). However, besides the widely studied feature of
non-locality (with or without entanglement \cite{nlwe}), quantum
theory exhibits also another form of non-classical correlations which
is quantified by the {\em quantum discord} \cite{ollzur}. Discord is
widely studied in the literature both as a resource for information
processing \cite{broterdiscomp} and as an ubiquitous feature in
quantum statistical mechanics \cite{terbrodmaxdaem,maxdaem}. However,
discord has not been explored yet in the framework of general
probabilistic theories.

The definition of quantum discord was originally motivated by the
analysis of states describing a quantum system and a pointer in a
measurement, at a time just after the unitary interaction of the
system with the pointer and the environment \cite{ollzur}. In the
particular situation where the environment broadcasts the measurement
outcome, the information carried by the pointer has strongly classical
features, which are not present in the general case of any bipartite
quantum state. Loosely speaking, a quantum state has null discord when
it resembles a bipartite pointer-system state in the peculiar
situation described above. On the other hand, the operational
interpretation of quantum discord is more obscure, and many subsequent
papers tackled this point by providing physical and informational
consequences of non-null discord
\cite{maxdaem,dattashajicaves,shabanilidar}, or by criticizing the
definition \cite{terbrodmaxdaem}. Information theoretically, discord
is interpreted as the amount of entanglement consumed in state merging
\cite{win}.

In this paper we introduce a definition of discord in general {\em
  causal} probabilistic theories \cite{long}, namely theories where
{\em no signalling form the future} holds. The definition reduces to a
geometric measure of discord \cite{dakvedbru} in the quantum case,
while null discord states coincide with the customary ones. We then
prove that in theories where the separability condition coincides with
the null discord condition, the set of states is simplicial, namely
all pure states are jointly perfectly discriminable. In this case,
only {\em entangled} non-null discord states may exist. If no
entangled states are allowed, then the theory is classical.
Consequently, there exist states with non-null discord in all causal
probabilistic theories apart from the classical one. Non-null discord
is not a quantum feature, but more generally a precise signature of
non-classicality of the theory. Thus, our result strengthens the
interpretation of non-null discord as non-classicality of
correlations.

The operational interpretation of discord of Ref. \cite{win} relies on
a definition in terms of mutual information, thus depending on the
notion of von Neumann entropy. However, there is no unique extension
of the von Neumann entropy for general probabilistic theories
\cite{barn,short,kimu}. Therefore we need an alternative definition,
relying on purely operational concepts. For this purpose, we will
first introduce a definition of null discord, extending a necessary
and sufficient condition stated in Ref.  \cite{ollzur}. We will then
define discord of a state $\rho$ in operational theories as the
minimum operational distance between $\rho$ and states with null
discord.

We briefly remind here the definitions of quantum discord introduced in
Refs.  \cite{ollzur}. Given a composite quantum system
$\rA\rB$ in state $\rho_{\rA\rB}$, the quantum mutual information
between $\rA$ and $\rB$ is defined as follows
\begin{equation}\label{mut1}
  {I}(\rA:\rB) = S(\rho_\rA) + S(\rho_{\rB}) - S(\rho_{\rA \rB}),
\end{equation}
where $S(\rho_\rX):=-\Tr[\rho_\rX\log\rho_\rX]$ is the Von Neumann
entropy of $\rho_\rX$, and $\rho_{\rA}$ and $\rho_\rB$ are the
marginal states $\Tr_\rB[\rho_{\rA\rB}]$ and $\Tr_\rA[\rho_{\rA\rB}]$,
respectively. The quantum mutual information can be considered as an
index of correlation. A second relevant measure of correlations can be
defined as follows. First, let us introduce a von Neumann POVM
$\mathbf \Pi^{\rA}=\{\Pi_i=|i\>\<i|\}_{i=1}^{d_\rA}$ for system $\rA$
with dimension $d_\rA$. Then, upon defining the conditional states
$\rho^{(i)}_{\rB} := \Tr_\rA [(\Pi_i\otimes I) \rho_{\rA\rB}(
\Pi_i\otimes I)]/p_i$, where $p_i:=\Tr[ (\Pi_i\otimes I)
\rho_{\rA\rB}]$, one can introduce the quantity
\begin{equation}\label{condent}
  {J}(\mathbf\Pi^\rA:\rB):=S(\rho_{\rB})- \sum_{i} p_i S(\rho_\rB^{(i)}).
\end{equation}
Quantum discord is then defined as \cite{ollzur}
\begin{equation}\label{disc}
  \mathcal{D}_{\mathbf \Pi^\rA}(\rho_{\rA \rB}) = {I}(\rA:\rB) - {J}(\mathbf\Pi^\rA:\rB).
\end{equation}
The dependence on the measurement $\mathbf\Pi^\rA$ is usually removed
by taking the minimum over possible von Neumann measurements. We
remark that quantum discord is an asymmetric quantity, due to the
definition of ${J}(\rA:\rB)$, and in general
$\mathcal{D}(\rho)\geq0$ for any $\rho$. In Ref.  \cite{ollzur} a
necessary and sufficient condition for null discord was introduced. We
now provide the condition as stated in Ref. \cite{dakvedbru}
\begin{theorem}\label{th:equc}
  A state $\rho$ of system $\rA\rB$ has null discord if and only if
  there exists a von Neumann measurement $\mathbf\Pi^\rA$ with
  $\Pi_k=|\psi_k\>\<\psi_k|$ on system $\rA$ such that
  \begin{equation}
    \sum_{k=1}^{d_\rA}(\Pi_k\otimes I_\rB)\rho(\Pi_k\otimes I_\rB)=\rho_{\rA\rB}.
    \label{eq:equicond}
  \end{equation}
\end{theorem}

The condition of theorem \ref{th:equc} will be taken as the definition
of null discord state for the purpose of generalizing the notion of
discord to the scenario of general probabilistic theories. We will now
briefly review the framework of operational probabilistic theories
introduced in Ref. \cite{purif}, and recently adopted for an
operational axiomatization of Quantum Theory \cite{long}.
\emph{Systems} and \emph{tests} are the primitive notions of an
operational theory.  A test represents one use of a \emph{physical
  device}, like a Stern-Gerlach magnet, a beam splitter, or a photon
counter. When the test is performed, it produces an \emph{outcome} $i$
in some set $\rX$. A test can then be viewed as the collection of all
events labeled by outcomes in the set $\rX$. Every test is labeled by
an \emph{input} and an \emph{output system}, respectively.  These
labels establish the rules for connecting two tests, namely tests
$\{\tA_i\}_{i\in\rX}$ and $\{\tB_j\}_{j\in\rY}$ can be connected in a
sequence $\{\tB_j\circ\tA_i\}_{(i,j)\in\rX\times\rY}$ only if the
output of the first test $\{\tA_i\}_{i\in\rX}$ is of the same type as
the input system of the second one $\{\tB_j\}_{j\in\rY}$. Systems are
denoted by capital letters, like $\rA, \rB, \rC,\dots$. We reserve the
letter $\rI$ for the \emph{trivial system}. A test with input or
output system $\rI$ is called {\em preparation test} or {\em
  observation test}, respectively.  Two systems $\rA$ and $\rB$ can be
composed in parallel, obtaining a third system $\rC:=\rA\rB$. Parallel
composition is commutative ($\rA\rB=\rB\rA$) and associative
($\rA(\rB\rC)=(\rA\rB)\rC$), and the trivial system acts as a unit
with respect to composition ($\rA\rI=\rI\rA=\rA$). An {\em operational
  theory} is specified by a collection of systems, closed under
composition, and by a collection of tests, closed under parallel and
sequential composition. An operational theory is \emph{probabilistic}
if every test $\{p_i\}_{i \in \rX}$ from the trivial system $\rI$ to
itself is a probability distribution over $\rX$, and both parallel and
sequential composition of two events from the trivial system to itself
are given by the product of probabilities: $p_i\otimes q_j= p_i \circ
q_j= p_i q_j$.

In a probabilistic theory, a preparation-event $\rho_i$ for system
$\rA$ defines a function $\hat \rho_i$ sending observation-events of
$\rA$ to probabilities: $\hat \rho_i: \Cntset (\rA) \to [0,1]$,
$a_j\mapsto a_j\circ\rho_i$.
Likewise, an observation-event $a_j$ defines a function $\hat a_j$
from preparation-events to probabilities $\hat a_j: \Stset (\rA) \to
[0,1]$, $ \rho_i\mapsto a_j\circ\rho_i$. Two observation-events
(preparation-events) are equivalent if they define the same function.
We will call the corresponding equivalence classes {\em states} ({\em
  effects}).
Since states (effects) are functions from effects (states) to
probabilities, one can take linear combinations of them. This defines
two real vector spaces $\Stset_\Reals (\rA)$ and $\Cntset_\Reals
(\rA)$, and here we restrict our attention to the case where such
spaces are finite dimensional. In this case, by construction one has
$D_\rA:=\dim (\Stset_\Reals (\rA)) = \dim (\Cntset_\Reals (\rA))$.

The scenario depicted up to now entails a wide variety of possible
theories, and we want now to restrict it as slightly as possible. In
particular, throughout the paper we will consider {\em causal}
theories \cite{mauro,purif}, which are defined as follows.

\begin{definition}[Causal theory] A theory is \emph{causal} if for
  every preparation-test $\{\rho_i\}_{i \in \rX}$ and every
  observation-test $\{a_j\}_{j \in \rY}$ on system $\rA$ the marginal
  probability $p_i := \sum_{j\in\rY} a_j\circ\rho_i$ is
  independent of the choice of the observation-test $\{a_j\}_{j \in
    \rY}$.  Precisely, if $\{a_j\}_{j \in \rY}$ and $\{b_k\}_{k \in
    \rZ}$ are two different observation-tests, then one has $\sum_{j
    \in \rY} a_j\circ\rho_i= \sum_{k \in \rZ} b_k\circ\rho_i$.
\end{definition}

Causal theories have a simple characterization given by the following
equivalent condition
\begin{lemma}[Characterization of causal theories]\label{lem:charcaus} A theory is causal if and only if
  for every system $\rA$ there is a unique deterministic effect
  $e_\rA$.
\end{lemma}

In this framework a separable state for a bipartite system $\rA\rB$ is
given by the following:

\begin{definition}
  A state $\rho$ of system $\rA\rB$ is separable if it is a
  convex combination factorized states, in formula
  \begin{equation}
    \rho = \sum_{i \in \rZ} p_i \rho_i \otimes \sigma_i
  \end{equation}
  where $\rho_i$ and $\sigma_i$ are states of systems $\rA$ and $\rB$,
  respectively, and $\{p_i\}_{i\in\rZ}$ is a probability distribution.
\end{definition}

Notice that we use the symbol $\otimes$ to denote the composition of
local states. However, in general causal theories without {\em local
  discriminability} the state space of a composite system is strictly
larger than the tensor product of the state spaces of the component
systems. However, separable states by definition lie in the subspace
defined by the tensor product of states of $\rA$ and states of $\rB$.

We now introduce the notion of {\em pure state}, which plays an
important role in the derivation of our results.
\begin{definition}[Pure and mixed states] 
  A state is {\em pure} if it cannot be written as a convex combination of
  other states. A state that is not pure is {\em mixed}.
\end{definition}

Finally, we need to introduce the notion of perfectly distinguishable
states.

\begin{definition}{\bf (Perfectly distinguishable states)}\label{def:distinguishable}
  The states $\{\rho_i\}_{i \in \rX} $ are \emph{perfectly
    distinguishable} if there is a test $\{a_i\}_{i \in \rX}$ such
  that $a_j\circ\rho_i = \delta_{ij}$. The test $\{a_i\}_{i \in
    \rX}$ is called \emph{discriminating test}.
\end{definition}

Now we have all the ingredients that are needed to export the notion
of null-discord state as provided by the condition of
Eq.~\eqref{eq:equicond} in the scenario of causal operational
probabilistic theories. First we will introduce the notion of
objective information, that can be viewed as an extension of the
notion of {\em element of reality} provided by Einstein, Podolski and
Rosen in their famous paper \cite{epr}. Notice however that we avoid
here any reference to the notion of real information and
reality.

\begin{definition}[Objective information]
  We say that a test $\{\tA_i\}_{i\in\rX}$ provides objective
  information about state $\rho$ if it fulfills the following
  requirements
  \begin{enumerate}
  \item The test $\{\tA_i\}_{i\in\rX}\in\Trnset(\rA)$ is repeatable, namely\\
    $\tA_i\circ\tA_j
    =\delta_{ij}\tA_i$.\label{prop:repe}
  \item The state $\rho$ is not disturbed by the test
    $\{\tA_i\}_{i\in\rX}$, namely $\tA\circ\rho=\rho$ for
    $\tA=\sum_{i\in\rX}\tA_i$.\label{prop:nondist}
  \end{enumerate}
  Equivalently, we will say that $\rho$ encodes objective information
  about the test $\{\tA_i\}_{i\in\rX}$
\end{definition}

This definition provides indeed the notion of a test that can extract
information from a system without disturbing its state, thus leaving
the same information accessible to further observers. As a consequence
of the definition, we can prove the following results.

\begin{lemma}
  If $\{\tA_i\}_{i\in\rX}$ provides objective information about the
  state $\rho$, then the states $\rho_i:=\tA_i\circ\rho/(e\circ\tA_i\circ\rho)$ are
  perfectly distinguishable by the test $a_i:=e\circ\tA_i$.
  \label{lem:objdisc}
\end{lemma}

\Proof Trivially follows from property \ref{prop:repe}.\qed

\begin{lemma}\label{lem:mixdiscr}
  If $\{\tA_i\}_{i\in\rX}$ provides objective information about the
  state $\rho$, then $\rho=\sum_{i\in\rX}p_i\rho_i$, where
  $p_i:=(e\circ\tA_i\circ\rho)$.
\end{lemma}

\Proof By the property \ref{prop:nondist} we have
$\rho=\tA\circ\rho=\sum_{i\in\rX}\tA_i\circ\rho=\sum_{i\in\rX}p_i\rho_i$.
Along with lemma \ref{lem:objdisc}, this proves the thesis.\qed

Finally, we introduce the following definition that accounts for those
situations where the objective information encoded in a state cannot
be further refined.
\begin{definition}
  A test $\{\tA_i\}_{i\in\rX}$ provides {\em complete} objective
  information about the state $\rho$ if it provides objective
  information about $\rho$ and the state $\rho_i$ is pure for all $i$.
\end{definition}

We will now define discord for operational probabilistic theories in
three steps.
\begin{enumerate}
\item We define null discord states.
\item We define the operational distance between two states of a
  generic probabilistic theory.
\item We define the discord $\mathcal{D}(\rho)$ of a bipartite state
  $\rho$ as the minimum of the distance between
  $\rho$ and the set of states with null discord.
\end{enumerate}

Let us now define null discord states as follows.
\begin{definition}[Null discord states]\label{def:zerodisc}
  In a causal operational probabilistic theory, a bipartite state
  $\rho$ has null discord if and only if it satisfies the following
  conditions
  \begin{enumerate}
  \item $\rho$ is separable, 
  \item there exists a test $\{\tA_k\}_{k\in\rX}$ on system $\rA$ that
    provides complete objective information about the state
    $e_\rB\circ\rho$, \label{cond:locinf} and such that $\{\tA_k
    \otimes \tI_{\rB}\}_{k\in\rX}$ provides objective information on
    $\rho$
  \end{enumerate}
\end{definition}

Notice that the notion of null discord states is not symmetric with
respect to the exchange of systems $\rA$ and $\rB$. In the following
we will follow the rule that null discord states encode objective
information on system $\rA$. We now state an equivalent condition for
null discord.
\begin{theorem}
  If the state $\rho$ has null discord, then it can be
  expressed as follows 
  \begin{equation}\label{eq:nulldi}
    \rho=\sum_{k\in\rX}q_{k}(\psi_k\otimes\sigma_k),
  \end{equation}
  where $\{\psi_k\}_{k\in\rX}$ is a set of jointly perfectly
  distinguishable pure states and $\{q_k\}_{k\in\rX}$ is a probability
  distribution.
\end{theorem}

\Proof Since $\rho$ is separable $\rho = \sum_{j\in\rY} p_j \rho_j
\otimes \tau_j$.  Moreover we have $\nu:=e_{\rB}\circ\rho=
\sum_{j\in\rY} p_j \rho_j$.  Since there exists a test
$\{\tA_k\}_{k\in\rX}$ that provides complete objective information on
$\nu$ we have $\tA_k\circ\nu
=\sum_{j\in\rY}p_j\tA_k\circ\rho_j = q_k \psi_k$, with $\psi_k$ pure
states. This means that
$\tA_k \circ\rho_j = p_{jk} \psi_k$ with $\sum_j p_j p_{jk} =: q_k$, namely
all vectors $\tA_k\circ\rho_j$ are parallel to each other and to the vector
$\psi_k$.  Since the test $\{\tA_k \otimes \tI\}_{k\in\rX}$ provides
objective information for $\rho$ we have $\sum_{k\in\rX}(
\tA_k \otimes \tI)\circ \rho = \sum_{j\in\rY}\sum_{k\in\rX} p_j
(\tA_k \circ\rho_j \otimes \tau_j) = \rho_{\rA\rB}$. Thus, exploiting the
fact that $\tA_k\circ\rho_j = p_{jk} \psi_k$, the latter expression becomes
\begin{equation}
  \rho_{\rA\rB} = \sum_{k\in\rX} q_{k} (\psi_k \otimes \sigma_k), 
\end{equation}
with $\{\psi_k\}_{k\in\rX}$ perfectly distinguishable states by
hypothesis, and $\sigma_k:=1/q_k\sum_{j\in\rY}p_jp_{jk}\tau_j$.
\qed

Let us now proceed to the operational definition of a distance between
states \cite{purif}. The operational distance between $\rho_0$ and
$\rho_1$ is defined through the minimum error probability
$p^{m}_{\mathrm{err}}$ in discrimination of $\rho_0$ and $\rho_1$
provided that their prior probability is $1/2$, namely
\begin{equation}
  \|\rho_1-\rho_0\|_{\mathrm{op}} := 1-2 p^{m}_{\mathrm{err}}.
\end{equation}

\vspace{0.2 cm} Let $\Omega_{\rA\rB}$ be the set of states of $\rA\rB$
with null discord. We can finally define the operational discord in a
generic probabilistic theory as follows.
\begin{definition}
  Given a probabilistic theory and a bipartite system $\rA\rB$, we
  define the discord $\mathcal{D}(\rho)$ of the state $\rho$ as
  follows
\begin{equation}
  \mathcal{D}(\rho):=\min_{\sigma \in \Omega_{\rA\rB}} \|\rho - \sigma\|_{\mathrm{op}}.
\end{equation}
\end{definition}
The present definition is hardly reducible to the standard notion of
discord in the quantum case. However, it is strictly related to a
geometric notion of discord proposed in \cite{dakvedbru}. Moreover,
for the purpose of the main results of the present paper, what matters
is the definition of null-discord states, which on the other hand
coincides with the standard one in the quantum case. We now prove the
main result of the present paper.
\begin{theorem}
  In a causal probabilistic theory where all separable states have
  null discord the set of normalized states for every system is a
  simplex.
  \label{th:simpl}
\end{theorem}
\Proof The proof consists in showing that equivalence of null discord
and separability in a causal probabilistic theory, implies that all
states of any system in the theory are convex combinations of the same
set of perfectly distinguishable states. Consider an arbitrary
separable state $\rho:=\sum_{i\in\rZ} p_i (\rho_i \otimes
\tau_i) $ with system $\rB$ equivalent to system $\rA$. By the
hypotheses of equivalence of separability and null discord, exploiting
Eq.~\eqref{eq:nulldi} we can write
\begin{equation}\label{equi}
  \rho=\sum_{i\in\rZ} p_i (\rho_i \otimes \tau_i) = \sum_{k\in\rX} q_{k} (\psi_k \otimes \sigma_k).
\end{equation}
In particular, we can consider states $\rho$ such that
$\{\rho_i\}_{i\in\rZ}$ and $\{\tau_i\}_{i\in\rZ}$ are complete sets of
linearly independent states of systems $\rA$ and $\rB$, respectively.
Now, by Eq.~\eqref{eq:nulldi}, if the test providing complete
objective information about the state $\nu:=e_\rB\circ\rho$ is
$\{\tA_k\}_{k\in\rX}$, then the observation-test $\{a_k\}_{k\in\rX}$
with $a_k:=e_\rA\circ\tA_k$ is such that $a_k\circ\psi_{k'} =
\delta_{kk'}$.  Applying $(a_{k}\otimes\tI_\rB)$ on state $\rho$, by
Eq.  (\ref{equi}) we have that $q_{k}\sigma_{k}= \sum_{i\in\rZ} p_i
(a_{k}\circ\rho_i)\tau_i, $ 
and substituting this expression into Eq.~(\ref{equi}) we have $
\sum_{i\in\rZ} p_i (\rho_i\otimes \tau_i)=
\sum_{i\in\rZ}p_i\sum_{k\in\rX} (a_k\circ\rho_i) (\psi_k\otimes
\tau_i) $.  Finally, by linear independence of $\tau_i$, for any $i$
we have $\rho_i=\sum_{k\in\rX} (a_k\circ\rho_i) \psi_k$.  Now, by
hypothesis of completeness of $\{\rho_i\}_{i\in\rZ}$, we can write any
state $\lambda$ of system $\rA$ as $\lambda = \sum_{i\in\rZ} c_i \rho_i$,
where $c_i$ are real numbers. Substituting the expansion of $\rho_i$
as a combination of states $\psi_k$ into the latter formula, we obtain
$\lambda= \sum_{k\in\rX} d_k\psi_k$, where we defined $d_l :=
\sum_{i\in\rZ} c_i (a_l\circ\rho_i)$.  Since $0\leq (a_k|\rho)=d_k$,
and $1=e_\rA\circ\lambda=\sum_{k\in\rZ}d_k$, we conclude that
$\{d_l\}_{l\in\rX}$ is a probability distribution. Hence we obtained
that any state $\lambda$ of system $\rA$ can be written as a convex
combination of the same set of perfectly distinguishable pure states
$\{\psi_k\}_{k\in\rX}$.$\blacksquare$

As a consequence of theorem \ref{th:simpl}, either the theory enjoys
local tomography \cite{mauro,purif} and then it is classical
probability theory, or it allows for entangled states---having
non-null discord---despite being simplicial. We can then conclude that
the only theory where no state has non-null discord is classical
probability theory.

In conclusion we introduced the notion of objective information in
causal theories, and used it to define null-discord states in the
general operational probabilistic framework. The notion of discord is
then introduced in terms of the minimum operational distance between a
given state and the set of null-discord states. These definitions
allowed us to prove that a theory where all separable states have null
discord must have simplicial state sets. Now, either the theory enjoys
local tomography and then it is classical, or it contains entangled
states, having non-null discord. As a consequence, the only theory
where no state has non-null discord is classical probability theory.
In view of this result, we can justify the widespread identification
of discord as a quantifier of non-classical correlations. Therefore,
discord is not at all a signature of quantumness, but one should
rather say that the absence of discord represents a singular feature
of classical probability theory among all causal probabilistic
theories.  Finally, we want to point out that the notion of objective
information introduced in this paper with the purpose of generalizing
the notion of {\em element of reality} of Ref.~\cite{epr} is a useful
tool in the context of operational probabilistic theories, with a
possible application to the extension of the notion of non-locality
without entanglement \cite{nlwe} in this framework.

\acknowledgments This work has been supported by EU FP7 programme
through the STREP project COQUIT. I thank G. M. D'Ariano for useful
discussion and advice.

\end{document}